\documentclass[12pt]{iopart}

\usepackage{iopams}  
\usepackage{epsfig}
\begin{document}

\comment[Molecular magnets]{Another dimension: investigations of molecular magnetism using
  muon-spin relaxation}

\author{Tom Lancaster}
\address{Durham University, Department of Physics, South Road, Durham,
  DH1 3LE
UK}
\ead{tom.lancaster@durham.ac.uk}
\author{Stephen J Blundell}
\address{University of Oxford, Clarendon Laboratory, Parks Road,
  Oxford, OX1 3PU, UK}
\author{Francis L Pratt}
\address{STFC ISIS Facility, Rutherford Appleton Laboratory, Chilton
  Didcot, OX11 0QX, UK}

\begin{abstract}
We review examples of muon-spin relaxation measurements on
molecule-based magnetic coordination polymers, classified by their
magnetic dimensionality. These include
the one-dimensional $s=1/2$ spin chain Cu(pyz)(NO$_{3}$)$_{2}$ 
and the two-dimensional $s=1/2$ layered material
[Cu(HF$_{2}$)(pyz)$_{2}$]BF$_{4}$. We also describe some of the more
exotic ground
states that may become accessible in the future given the ability to tune the
interaction strengths of our materials through crystal engineering. 
\end{abstract}

\maketitle


\section{Molecule-based magnets}

Molecule-based magnets are systems whose principal structural building blocks
are organic molecules \cite{molmagreview_1,molmagreview_2}. 
We understand the microscopic origin of the  magnetism that
arises from such materials in terms of paramagnetic
spin centres linked via superexchange pathways. The magnetic centres themselves
 may be atoms (such as $s=1/2$ Cu$^{2+}$) or
uncompensated spins on the molecules (as in the case of
radical magnets) and the superexchange pathways will frequently pass
through the molecular groups. The great promise of choosing to make
our magnets out of molecules is {\it tunability}, that is, the possibility
of manipulating the carbon chemistry of the molecules to form
structures which promote magnetic exchange in certain geometries and
to manipulate the strength of the coupling between the magnetic units. 

Research on molecular magnetic systems relies on a confluence of experimental and
theoretical physics and chemistry. Chemistry is vital in the synthesis
of new materials with the desired structural properties which incorporate
the necessary magnetic ingredients; experiment allows us to probe the
magnetism and we hope that theory can explain it. In this Comment we will discuss some examples of the use
and successes 
of muon-spin relaxation ($\mu^{+}$SR) \cite{steve} as an experimental technique in
this area. We will classify the systems that we discuss in terms of
their
magnetic dimensionality, that is, the number of dimensions in which
the coupling between magnetic units is strong.  
The explanation of the physics of such systems is
usually given in terms of low-energy models of exchange-coupled spins
\cite{tsvelik,sachdev,giamarchi} whose
long-wavelength properties have been elucidated using field-theoretic techniques and, in
particular, the renormalization group (RG) \cite{wilson,brezin}. In presenting some possible
future directions for muon spectroscopy in this field we will describe
some of the theoretical models which are currently motivating the
synthesis of new materials. 

The subject of molecular magnetism has been reviewed thoroughly
elsewhere \cite{molmagreview_1,molmagreview_2}, as has early, and
more recent,
work using $\mu^{+}$SR \cite{mu_review1,mu_review2,mu_review3,mu_review4,mu_review5}. 
However, it is worth mentioning some of the notable themes from the
early muon work, which 
include: (i) the
observation of long-range magnetic order  (LRO), such as in the purely organic material 
$\beta$-phase {\it p}-NPNN, where oscillations in the muon asymmetry are
found below 0.7~K \cite{npnn}; (ii) the observation of dynamics in molecular
nanomagnets such as the single molecule magnet Mn$_{12}$Ac
\cite{lascialfari} and (iii) the
possibility of observing collective magnetic behaviour such as the
spin-crossover transition \cite{mu_review1}. These investigations play to the
strengths of $\mu^{+}$SR in that they require a probe sensitive to
 small magnetic moments and their dynamics and the possibility of
 distinguishing between uniform magnetic effects and minority,
 strongly magnetic phases. 

The Comment is structured as follows: in section~\ref{1d} we will
discuss the theoretical prediction and realization of one-dimensional
magnetism from molecular chains. In section~\ref{2d} we turn to
layered two-dimensional systems
and in section~\ref{dimsel} to the recently realised possibility of switching between
dimensionalities. Finally we discuss 
some of the possible future directions for this field in section~\ref{frontiers}.

\section{Magnetism in one dimension}
\label{1d}

Much recent work on molecular magnets
using $\mu^{+}$SR has concerned
materials whose low-energy physics may be thought of as arising from 
interactions which are constrained to act in less than three spatial
dimensions. Such physics usually achieved by forming an array of
transition metal ions, such as $s=1/2$ Cu$^{2+}$ and linking these
using one or more molecular ligands in such a way as to promote strong
interactions between the magnetic ions only along particular directions and
weaker ones along others. 

\begin{figure}
\begin{center}
\epsfig{file=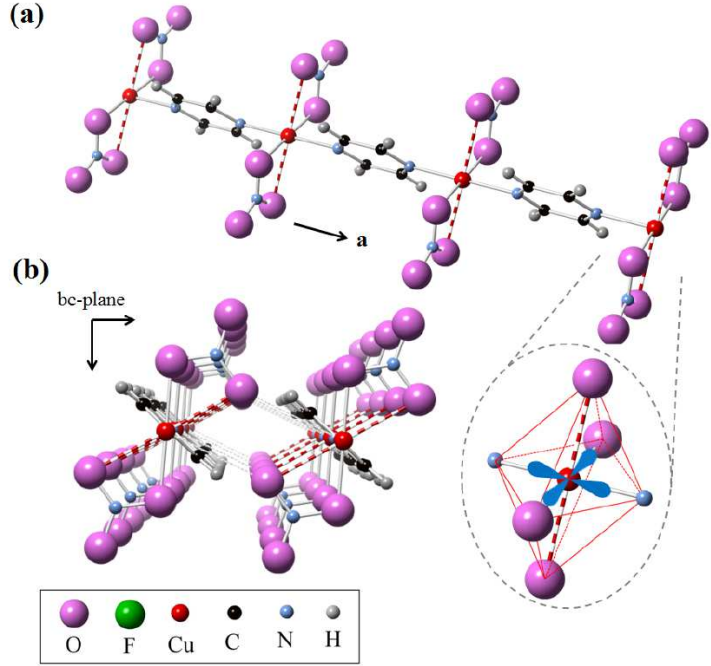,width=10cm}
\caption{The structure of Cu(pyz)(NO$_{3}$)$_{2}$. (a) The Cu-pyz-Cu
  chains. (b) The chain packing viewed along the
  $a$-direction. Figure taken from Ref.~\cite{pressure}.\label{cupzno3}}
\end{center}
\end{figure}

As an example, we will consider a material whose structure promotes
strong, antiferromagnetic superexchange along
one-dimensional, parallel, linear chains, and weaker exchange in the other two
perpendicular directions. Such a
material may be constructed if we link
Cu$^{2+}$ ions along one direction using pyrazine (pyz $\equiv$ C$_{4}$H$_{4}$N$_{2}$) ligands and separate
the
chains using NO$_{3}$ groups. The resulting material formed from this 
recipe is Cu(pyz)(NO$_{3}$)$_{2}$ and is shown in figure~\ref{cupzno3}. 

To understand the physics of such materials we seek to describe them using
simple spin models.
To a good first approximation the crystal field surrounding the
Cu$^{2+}$ ions does not cause the spin to favour any particular spatial
direction, resulting in a so-called Heisenberg spin \cite{steve_book}.  
The resulting one-dimensional ($d=1$) quantum Heisenberg antiferromagnet
(1DQHAF) may be described by a model Hamiltonian \cite{tsvelik,sachdev,note}
\begin{equation}
\hat{H} = J \sum_{\langle ij\rangle}
\hat{\boldsymbol{S}}_{i}\cdot\hat{\boldsymbol{S}}_{j}, 
\label{model1}
\end{equation}
where $i$ and $j$ label spins along the same chain and the sum is
over unique nearest neighbours bonds only. For materials such as
Cu(pyz)(NO$_{3}$)$_{2}$ the parameter $J$ may be determined from the
result of magnetic susceptibility or pulsed field magnetization (see
below). 

\begin{figure}
\begin{center}
\epsfig{file=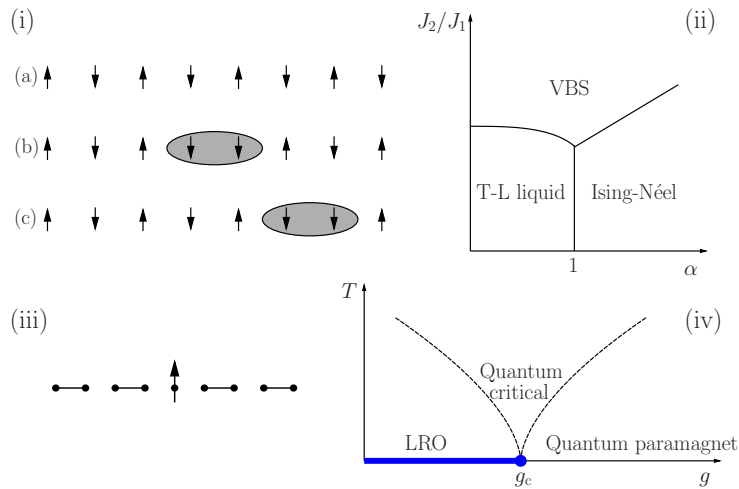,width=10cm}
\caption{(i) The spinon excitation in a $d=1$ chain. 
 (ii) The phase diagram of a $d=1$ chain system with nearest
 neighbour coupling $J_{1}$, next-nearest neighbour coupling $J_{2}$ and
anisotropy parameter $\alpha$ [VBS=valence bond solid; T-L liquid=Tomonaga
 Luttinger liquid  (see Section~\ref{frontiers}).]
(iii) The valence bond solid with a single $s=1/2$ excitation. 
 (iv) The phase diagram of a $d=2$ system showing a quantum critical
 point at a coupling $g_{\mathrm{c}}$ and a quantum critical region at
 elevated temperatures. \label{theory}}
\end{center}
\end{figure}

The model Hamiltonian in eqn~\ref{model1} is very
interesting and has motivated a wide range of research \cite{tsvelik,sachdev}. 
It possesses a continuous $O(3)$ symmetry (here the global
rotation of all of the spins through an arbitrary angle). In common with
other $d=1$ models with finite range interactions
there will be no LRO for $T > 0$. (For $d=1$ this is
easy to see: pass the message ``line-up'' along the
chain in the presence of a defect (or a fluctuation) and the message
is lost at the weak link: since each spin is only able to communicate with its
neighbours the message cannot pass around the weak link). 
Although this problem may be solved exactly using the famous Bethe
ansatz \cite{giamarchi} the solution is rather unwieldy and many of our insights
arise from the predictions of a continuum model which, as long as the
long-range spin-spin
coupling is not too large (see section~\ref{frontiers}), results
from
mapping the problem onto a $d=1$ model of spinless fermions and then using a 
technique known as bosonization. 
The predicted state is known as a Tomonaga-Luttinger (T-L) liquid
\cite{sachdev,giamarchi} and its 
 excitation spectrum is one of gapless, linearly dispersing,
 particle-like excitations. Some physical insight into the nature of
 these excitations comes from figure~\ref{theory}(i). An
 antiferromagnetically correlated region [as shown in (a)] may be excited if we flip
 a section of the spins, as has been done after the fourth spin in
 (b). This creates a spin 1/2 excitation, known as a {\it spinon}
 [shown shaded in figure~\ref{theory}(i)],
 which may move by hopping by two lattice spacings at a time [shown in (c)]. 

\begin{figure}
\begin{center}
\epsfig{file=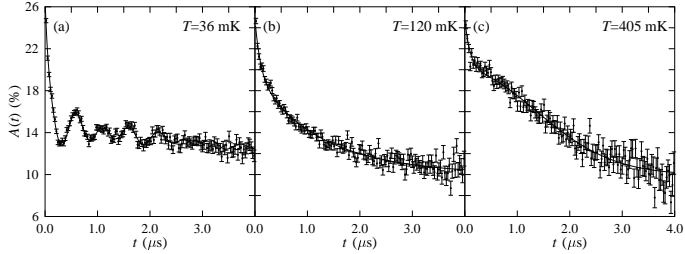,width=10cm}
\caption{Zero field muon-spin relaxation data measured on
  Cu(pyz)(NO$_{3}$)$_{2}$, showing long-range magnetic order below
  $T_{\mathrm{N}}=107$~mK \cite{cupyzno3}. \label{cupzno3_data}.}
\end{center}
\end{figure}

The idealisation that leads to the T-L model has ignored many
interactions in Cu(pyz)(NO$_{3}$)$_{2}$ and one of these proves particularly relevant: the
interaction between chains parametrised by an exchange interaction
$J_{\perp}$. This changes the Hamiltonian to
\begin{equation}
\hat{H} = J \sum_{\langle ij\rangle}
\hat{\boldsymbol{S}}_{i}\cdot\hat{\boldsymbol{S}}_{j}
+J_{\perp} \sum_{\langle ij'\rangle}
\hat{\boldsymbol{S}}_{i}\cdot\hat{\boldsymbol{S}}_{j'}, 
\label{model2}
\end{equation}
where $i$ and $j'$ label spins on adjacent chains. 
In order to probe $d=1$ behavior a system described by this model we would need to
work in a temperature regime 
where $J_{\perp} \ll  T \ll J$. Once we start to approach a regime
where the characteristic energy of the thermal fluctuations allows
detail on the level of
$J_{\perp}$ to be resolved, we should expect that the three-dimensional nature of the
material  becomes evident and it will 
magnetically order. However, the strongly anisotropic nature of the
exchange ($J_{\perp}/J\ll 1$) will have consequences on the ordered
state that arises \cite{sengupta1} and the ordering temperature $T_{\mathrm{N}}$ is driven
downwards by the one-dimensional nature of the system compared to a
three-dimensional one.  
The ordered magnetic moments will also be
renormalized, making them potentially very small, and difficult to
observe with magnetic susceptibility. Finite, but possibly
very long, correlation lengths may exist above the ordering temperature, 
reducing the entropy in the system. Consequently, the entropy change
on ordering will be much reduced compared to a more isotropic system and
may prevent specific heat measurements from detecting a
transition. Add the difficulty of deuterating all of the
hydrogen-containing groups in the system (and the possibility of
deuteration changing the delicate superexchange pathways \cite{goddard_deuterate}) and neutron
diffraction also becomes difficult. 
It is here that $\mu^{+}$SR has proven very effective in detecting 
magnetic transitions in reduced dimensions that have proven
difficult to observe using more conventional experimental methods.
Before the $\mu^{+}$SR study, no magnetic order had been observed in Cu(pyz)(NO$_{3}$)$_{2}$ despite
specific heat measurements made down to 70~mK. However, the clear
zero-field precession signal from $\mu^{+}$SR \cite{cupyzno3} shows unambiguous
evidence for LRO below a transition temperature of
$T_{\mathrm{N}}=0.107$~K as shown in figure~\ref{cupzno3_data}. 

A useful figure of merit in comparing the success with which a material
realizes one-dimensionality is $T_{\mathrm{N}}/J$, since this
quantity should be zero in the ideal case and of order and close to unity for an
isotropic material. This quantity may also be
used to estimate $J_{\perp}$, which is the parameter of interest in the
Hamiltonian in eqn~\ref{model2}. One method of doing this is has been developed from
Quantum Monte Carlo computations \cite{yasuda}, whose results may be approximated 
using the expression
\begin{equation}
|J_{\perp}|/k_{\mathrm{B}} =\frac{T_{\mathrm{N}}}{
4 c \sqrt{  
\ln\left( 
\frac{a|J|}{\mathrm{k}_{\mathrm{B}}T_{\mathrm{N}}}
\right)
+
\frac{1}{2}\ln \ln
\left(
\frac{a |J|}{\mathrm{k}_{\mathrm{B}}T_{\mathrm{N}}}
\right)
}}, 
\end{equation}
with $a=2.6$ and $c=0.233$. 
This yields a value of $|J_{\perp}/J| = 4.4\times 10^{-3}$ for
Cu(pyz)(NO$_{3}$)$_{2}$, demonstrating that it is indeed a very well
isolated approximation of a $d=1$ magnet. 
 In addition, coupled chain mean field theory \cite{schulz} predicts an ordered
moment of 0.05~$\mu_{\mathrm{B}}$. 

The application of a magnetic field to our $d=1$ system acts to
polarize the spins. If we imagine doing this at $T=0$ then there will
be a point at a  field $\mu_{\mathrm{B}} B \gg J$ where all of the
spins will be aligned. This aligned state is separated from the
partially polarised antiferromagnet by a quantum critical point (QCP) \cite{sachdev} at a
field $B_{\mathrm{c}} = 2 J/ g \mu_{\mathrm{B}}$ (we neglect
interchain interactions here). One advantage of
investigating molecular magnetic systems such as these is that the
exchange strength, and hence the QCP, is typically of order 
$J \approx 10$~K, translating as $B_{\mathrm{c}}\approx
10$~T. Although such fields are currently outside the range accessible
via $\mu^{+}$SR they are accessible using pulsed magnetic fields and
the combination of high-field magnetization and muon spectroscopy
\cite{paul_njp}
has
proved valuable in fully characterising new materials in recent
years. These characteristic energy scales are orders of magnitude
smaller than the corresponding ones for inorganic spin chains
such as Sr$_{2}$CuO$_{3}$, whose QCP occurs at an experimentally 
inaccessible field around 200 times
larger than that of Cu(pyz)(NO$_{3}$)$_{2}$. 

An even more ideal example of this physics is found in the radical
system DEOCC-TCNQF$_{4}$ \cite{deocc}. In this case no LRO transition is observed
using $\mu^{+}$SR down to 20~mK, constraining $|J_{\perp}/J| <6 \times 10^{-5} $, making this the
most successful realization of a 1DQHAFM yet reported. In this case, the weak
relaxation that is observed via $\mu^{+}$SR allows the investigation
of the magnetic fluctuation spectrum. 
Whether the excitations in 1DQHAFs are ballistic or
diffusive remains 
a key question in the physics of spin chains 
\cite{sirka,steinigeweg}
and one that can be addressed with LF $\mu^{+}$SR measurements.
Field-dependent studies can be used to distinguish
between these types of spin transport, since their
spin autocorrelation functions have different associated
spectral densities $f(\omega)$ which vary as $ \omega^{-\frac{1}{2}}$
for $d=1$ diffusive transport and
$f(\omega) \propto \ln(J/\omega)$ for ballistic motion. 
In DEOCC-TCNQF$_{4}$ diffusive transport was identified. This
contrasts with the inorganic
 1DQHAF Rb$_{4}$Cu(MoO$_{4}$)$_{3}$ where ballistic transport was
shown to dominate the muon relaxation \cite{cumo}. The details and
general rules of the $d=1$ transport are complicated and this
continues to be an area exciting much interest and new research.

\section{Two dimensions}
\label{2d}

The model of the two-dimensional square lattice quantum Heisenberg
antiferromagnet (2DSLQHA) has received significant attention as
another apparently simple system where the influence of dimensionality leads to the
emergence of an unusual ground state and excitation spectrum \cite{manousakis}.
It is also, of course, the low-energy model that describes the Mott insulating 
parent phase of the cuprate superconductors. 
The 2DSLQHA shows no magnetic order for $T>0$, owing to the
Coleman-Mermin-Wagner theorem \cite{mermin,coleman2}, but there is good evidence that 
it orders at $T=0$ with a much renormalized ordered moment of around
0.62~$\mu_{\mathrm{B}}$ \cite{manousakis}.  
The excitations of the model at $T=0$ are gapless spin wave
excitations with a linear dispersion \cite{tsvelik,sachdev,manousakis}. 

\begin{figure}
\begin{center}
\epsfig{file=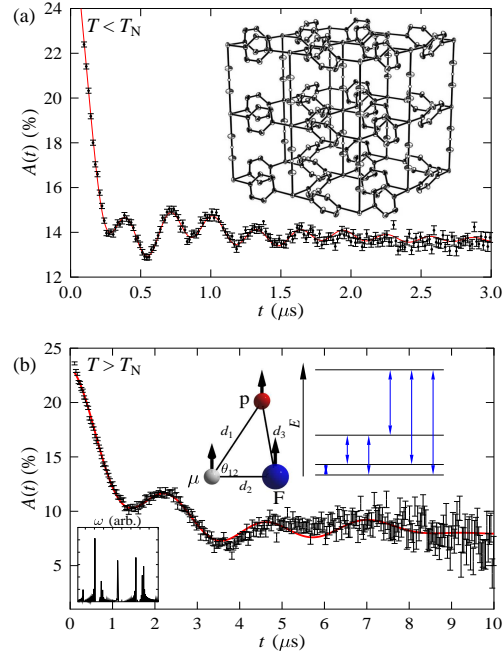,width=7cm}
\caption{Zero field $\mu^{+}$SR data measured on [CuHF$_{2}$(pyz)$_{2}$](BF$_{4}$). (a) Zero
  field
oscillations below $T_{\mathrm{N}}=1.54$~K allow us to identify 
LRO and probe the critical behaviour. \cite{manson} (b) For $T>T_{\mathrm{N}}$
we observe muon-fluorine dipole-dipole oscillations that allow us to
deduce muon sites \cite{steele,fmuf}. \label{2dfig}}
\end{center}
\end{figure}

Of the materials recently synthesised which show two-dimensional ($d=2$)
magnetic properties, one of the most successful classes is based on
two-dimensional square layers of Cu$^{2+}$ linked with pyrazine
groups to form [Cu(pyz)$_{2}$]$^{2+}$ layers with an intralayer exchange coupling $J$. 
These square layers may be linked with HF$^{2-}$ ions to form pseudocubic
arrays of [CuHF$_{2}$(pyz)$_{2}$]$X$ [shown inset in
Fig.~\ref{2dfig}(a) \cite{manson}], where $X$ is an anion that sits
in the centre of the cube and provides charge neutrality. 
As in the $d=1$ case, we are able
to identify magnetic order in this system at low temperature and
assess the degree to which it may be described as two-dimensional \cite{steele}. 
As an example we may consider [CuHF$_{2}$(pyz)$_{2}$]BF$_{4}$
(Fig.~\ref{2dfig}) which magnetically orders at $T=1.54$~K \cite{manson}. If we attribute
the ordering to an interlayer coupling $J_{\perp}$ we
may use the results of quantum Monte Carlo calculations which, in this
case, predict \cite{yasuda}
that 
\begin{equation}
|J_{\perp}/J|= \exp\left( 
b-\frac{4\pi\rho_{\mathrm{s}}}{T_{\mathrm{N}}},
\right),
\end{equation}
where $\rho_{\mathrm{s}} = 0.183 J$ is the spin stiffness and the
parameter $b=2.43$. For [CuHF$_{2}$(pyz)$_{2}$]BF$_{4}$ we find
$J=6.3$~K and $|J_{\perp}/J|=9\times 10^{-4}$, making this material
a highly anisotropic $d=2$ antiferromagnet. 

In addition, the fact that these systems contain fluorine leads to oscillations at
temperatures above $T_{\mathrm{N}}$ attributable to F-$\mu^{+}$
dipole-dipole coupling [Figure~\ref{2dfig}(b)]. These are useful here as they allow us to 
work out where the muons giving rise to these signals are localised with
respect to the fluorine nuclei in these materials \cite{fmuf}. It is
worth noting however that in complex materials such as these there are
often several classes of muon sites, not all of which may be accounted
for by those muons giving F-$\mu^{+}$ oscillations. 

Although interlayer coupling in a 2DQHAF again provides a means for magnetic order
to occur at non-zero temperatures there is evidence that in this
system a degree of  $g$-factor anisotropy causes the Cu$^{2+}$ ions
to take on a degree of $XY$-like behavior \cite{xiao}. This is of interest as the
pure $XY$-model allows the occurrence of a new sort of excitation: the
{\it vortex}. A single vortex is unstable alone, and in the absence of an
external electromagnetic field, we may only stabilise vortices at low
temperature by binding them with antivortices. The transition
between bound vortex-antivortex pairs and single vortices stabilised
by thermal fluctuations at high temperature was originally identified
by Kosterlitz and Thouless and by Berezinskii for an ideal system, but has never been
observed in a bulk magnet \cite{berezinskii,kosterlitz,bramwell}. However, the physics that those authors
identified may have an effect even when only a small $XY$-like
component is present. In this case we expect that, while at high
temperature
the spins will be Heisenberg-like, as the temperature is lowered to a
scale where the $XY$-like component is resolvable it becomes
energetically favourable for the spins to tip into the $XY$-plane. On
further
decrease of temperature the possibility of a vortex-binding transition
greatly increases the correlation length, at which point the
interlayer coupling $J_{\perp}$ causes the system to undergo a
transition to LRO \cite{xiao}. 

\begin{figure}
\begin{center}
\epsfig{file=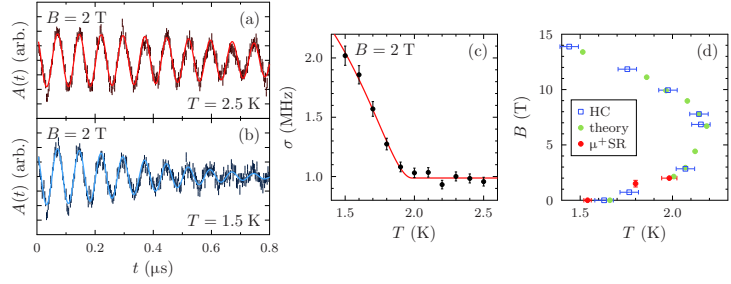,width=10cm}
\caption{(a,b) Transverse field $\mu^{+}$SR measurements on
  Cu(HF$_{2}$)(pyz)$_{2}$]BF$_{4}$. (c) The relaxation rate allow us to extract the
ordering temperature in field.  (d) The result is a non-monotonic phase
boundary caused by a small $XY$-like component in the 
spin Hamiltonian of the system \cite{steele}.
\label{nonmonotonic}}
\end{center}
\end{figure}

One experimental consequence of $XY$-like behaviour 
is that the phase boundary in applied magnetic field takes on a non-monotonic 
form as shown in figure~\ref{nonmonotonic} \cite{sengupta2,steele}. 
The application of the field makes it energetically favourable for the
spins to tip into the $XY$ plane, effectively reducing the
dimensionality. It then has a dual effect on the spins by
suppressing the amplitude of the order parameter by forcing spins
along a particular direction and also reducing
the phase space for phase fluctuations from a sphere to a circle. 
At low fields the magnetic ordering transition is controlled by phase fluctuations and at
higher field by amplitude fluctuations leading to the non-monotonic
phase boundary that is observed in experiment \cite{sengupta2}. 
We have been able to measure the phase boundary in field using both LF
$\mu^{+}$SR and transverse field  (TF) $\mu^{+}$SR \cite{steele} as
shown for [Cu(HF$_{2}$)(pyz)$_{2}$]BF$_{4}$ in
 figure~\ref{nonmonotonic}. 
It has been argued \cite{sengupta2} that the energy scales for this
physics are controlled by the Kosterlitz-Thouless physics described
 above. It is probable that the physics
of vortices, along more complicated topological objects such as merons
 and skyrmions, will be an area of future interest in this field.

\section{Switching the dimensionality}
\label{dimsel}

We might ask if it is possible to gain control over the dimensionality of a
system.  There have been two recent advances that suggest that this is
a fruitful area for future work: the first a synthetic route; the
second using pressure as a means of creating a dimensionality switch. 

The first route \cite{goddard1} relies on varying the proportions
of reactants in the material synthesis, which makes it possible
 to create two distinct, but closely related materials, one two-dimensional; the
other one-dimensional. 
The coordination polymer [Cu(pyz)$_{2}$(pyO)$_{2}$](PF$_{6}$)$_{2}$ (where pyO denotes
pyridine-N-oxide=C$_{5}$H$_{5}$NO) is another $d=2$ square lattice
material based on [Cu(pyz)$_{2}$]$^{2+}$ layers, this time with pyO
groups protruding from them. 
By changing the proportion of pyz and pyO molecules in the preparation
from 3:1 to 2:1 
the water-containing material
[Cu(pyz)(pyO)(H$_{2}$O)$_{2}$](PF$_{6}$)$_{2}$ is produced. 
Here pyz ligands link Cu atoms along the $b$-axis only, forming
highly one-dimensional chains. 
Crucially, the strong nearest-neighbour exchange interaction maintains a value of $J\approx 8$~K in both
cases. 

\begin{figure}
\begin{center}
\epsfig{file=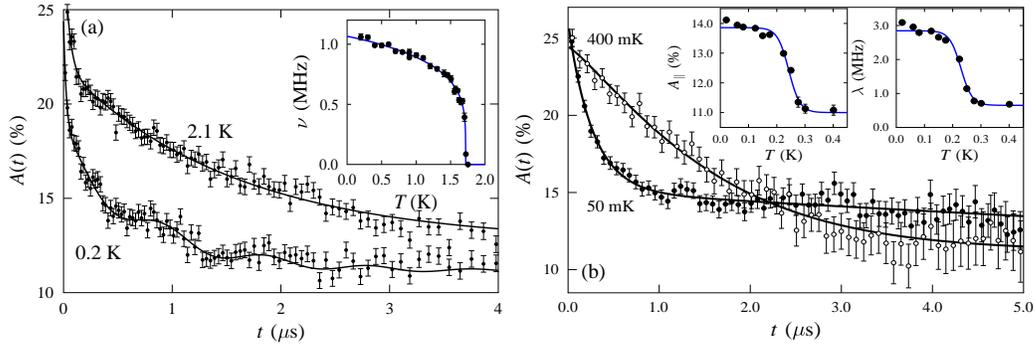,width=14cm}
\caption{Dimensionality switching: $\mu^{+}$SR allows us to probe 
magnetic order in (a)
$d=2$ [Cu(pyz)$_{2}$(pyO)$_{2}$](PF$_{6}$)$_{2}$ 
and (b) $d=1$
Cu(pyz)(pyO)(H$_{2}$O)$_{2}$](PF$_{6}$)$_{2}$. \cite{goddard1}
\label{fig:dimsel}}
\end{center}
\end{figure}

These structures suggest $d=2$ and $d=1$ magnetic behaviour,
which is indeed confirmed by
a combination of high-field magnetometry and muon-spin relaxation \cite{goddard1}. 
The results of the $\mu^{+}$SR measurements are shown in figure~\ref{fig:dimsel}
where we are able to establish where each material undergoes a
transition to LRO. The ordering temperature and magnetic moment size
of the $d=2$ material are both larger than in the analogous $d=1$
material, which reflects the expected increase in the influence of
fluctuations for $d=1$ compared to $d=2$. We conclude, in this case,
that to turn a material from $d=2$ to $d=1$, one must just add water!

The second method of controlling dimensionality
involves using pressure as a switch \cite{pressure}. 
The material CuF$_{2}$(H$_{2}$O)$_{2}$pyz \cite{cuf2} is based on a chain-like
structure but, rather counter-intuitively, it is a good
realization of a 2DQHAFM. The reason for this is that the
chains arrange themselves in such a way as to form square layers of
Cu$^{2+}$ ions with superexchange pathways provided by hydrogen
bonds.  Crucially the unfilled $d_{x^{2}-y^{2}}$ orbital lies in the plane of
the square layers, providing strong two-dimensional exchange 
with only weak coupling along the chains. 


Applying hydrostatic pressure causes a dramatic change in behaviour
above around 10~kbar \cite{pressure}. Here the magnetization shifts from that typical
of a $d=2$ system to that of a $d=1$ magnet. $\mu^{+}$SR has been
carried out under pressure and reveals the change in $T_{\mathrm{N}}$ as expected from a change from $d=2$ to $d=1$. 
The microscopic explanation for this quantum phase transition is suggested by a
combination of x-ray diffraction and electron spin resonance \cite{halder,prescimone}. The
application
of large hydrostatic pressures causes a small structural  distortion
which, in turn,  causes the Jahn-Teller axis of the Cu$^{2+}$ ions to
switch directions,
resulting in the $d_{x^{2}-y^{2}}$ orbital shifting from linking Cu
atoms in hydrogen bonded [Cu(pyz)$_{2}$] layers to point along the
 Cu-pyz-Cu chains. 

It is possible that such a switching of magnetic dimensionality
will be found, and ultimately engineered, in a variety of magnetic
coordination polymers. The fundamental change in the character of a
system in going from $d=2$ to $d=1$ is such that this switching feature could
conceivably find a use in future quantum technologies.

\section{Frontiers: a zoology of ground states}
\label{frontiers}

We now turn to a possible future direction for the engineering of
coordination polymers. We hope that it will soon be possible to
produce materials that realise still  more complicated models of
low-dimensional magnetic behaviour. These would allow us to realise a large
number of new and potentially exotic ground states. Here we briefly
review the predictions for some of the quantum states that may be
achieved if we are able to tune the interaction strength of our
$s=1/2$ spins. For the full story we refer you to some of the
excellent books on this subject \cite{tsvelik,sachdev,giamarchi}.

\subsection{$d=1$: Tomonaga-Luttinger, Valence bond solid and Ising-N\'{e}el states}

When describing the 1DQHAFM in section~\ref{1d} we claimed that the disordered
T-L ground state with its gapless linear excitation spectrum was reached if the coupling was
not too large. More precisely, we mean that that the antiferromagnetic
nearest neighbour
exchange coupling $J_{1}$
is dominant and the antiferromagnetic  next-nearest neighbour
exchange $J_{2}$ is small in comparison.
 However, if the ratio of the couplings exceeds a critical
 value $J_{2}/J_{2}>J_{\mathrm{c}}$
then the system adopts another ground state: the spin-Peierls 
or (in more modern terminology) the valence bond solid (VBS) state, illustrated
in figure~\ref{theory}(iii). The VBS ground state involves the spins
staggering their bond exchange energy, essentially forming
antiferromagnetically coupled dimers
along the chain. There are two ways of doing this along any
chain  (since a particular site may form a dimer with its
neighbour to the left, or to the right) and so the ground state that is realised 
must break translational symmetry. 
An antiferromagnetically coupled dimer has a singlet ground
state, resulting in a total $S=0$ ground state for the VBS. 
The excitations of the VBS phase are lone spins sat between the valence
bonds. These are spin $s=1/2$ excitations which, as in the T-L
case are  also known as
spinons but here there is an energy gap $\Delta$ which must be exceeded to
create such an excitation. In fact, 
the energy required to excite a spinon of momentum $\boldsymbol{p}$ is given by
$E_{\boldsymbol{p}} = \left(
  \Delta^{2}+v_{\mathrm{F}}^{2}\boldsymbol{p}^{2}\right)^{\frac{1}{2}}$,
where $v_{\mathrm{F}}$ is the Fermi energy of the system. 

Although we have hitherto concentrated only on low-energy models
which include purely spin degrees of freedom,
the realization of the VBS state also
requires the inclusion of magnetoelastic coupling (and it is in this context
that it is usually called a spin-Peierls state).
  Above the spin-Peierls transition temperature
$T_{\rm SP}$, we have a uniform antiferromagnetic next-neighbour
exchange in each chain; below $T_{\rm SP}$ there is an elastic
distortion resulting in the  dimerization described abve, and hence two, unequal,
alternating exchange constants.  The alternating chain possesses an
energy gap between the singlet ground state and the lowest lying band
of triplet excited states which closes up above $T_{\rm SP}$.  The
transition temperatures may be related to the relevant coupling
constants; whereas the conventional Peierls distortion is expected at
a temperature $T_{\rm P} \sim (E_{\rm F}/k_{\rm B})\exp(-1/\Lambda)$,
where $E_{\rm F}$ is the Fermi energy of the system and $\Lambda$ is
the electron-phonon coupling constant, the spin-Peierls transition is
expected at $T_{\rm sP} \sim (J/k_{\rm B})\exp(-1/\Lambda)$.  $J \ll
E_{\rm F}$ and hence $T_{\rm sP} \ll T_{\rm P}$ \cite{brayrev}.  

One example system is the organic salt MEM(TCNQ)$_2$, which consists
of one-dimensional stacks of planar TCNQ molecules, each of which has
a charge of $-{1 \over 2}e$ associated with it. Adjacent stacks are
separated by arrangements of MEM molecules, each of which possess a
localized charge of $+e$.  It undergoes two structural
distortions. The first, which occurs at 335~K, is a conventional
Peierls transition in which the TCNQ chains dimerize.  This results in
a change from metallic to insulating behaviour as a single electronic
charge becomes localized on each TCNQ dimer; the single spin on each
dimer couples antiferromagnetically to its neighbours. This phase
persists down to the spin-Peierls transition at 18~K, where a further
dimerization of the TCNQ stacks takes place (this is a tetramerization
of the original chain).  $\mu$SR studies indicate a slowing down of
the electronic fluctuations resulting from the opening of the
spin-Peierls gap in the magnetic excitation spectrum as the
temperature is lowered below $T_{\rm sP}$ \cite{mem}.  At the very
lowest temperatures the electronic spin fluctuations freeze out and
the muon-spin depolarization is dominated by persistent slow
fluctuations which can be ascribed to a defect-spin system
\cite{brendonmem2}.  It has been speculated that the muon spin might
locally create a spin defect by breaking a singlet pair, giving rise
to the relaxation ascribed to defect spins and which is only revealed
when the other sources of relaxation have frozen out
\cite{brendonmem2}.  These results have now been extended to the
related material DEM(TCNQ)$_2$ \cite{brazil}.  Other examples of
spin-Peierls systems include other organic systems such as
TTF-CuS$_4$C$_4$(CF$_3$)$_4$ ($T_{\mathrm{sP}}=12$~K) \cite{BDT} and
TTF-AuS$_4$C$_4$(CF$_3$)$_4$ ($T_{\mathrm{sP}}=2$~K) \cite{BDT,BDT2}.  A common
feature of such materials is flat organic molecules in columnar
stacks.  The large interchain separation and weak van der Waals
intermolecular interactions favour the dominance of magnetoelastic
effects over interchain ordering.  In contrast the chains in
corresponding inorganic materials, such as copper chain compounds, are
quite rigid due to the ionic bonding and only a single example of an
inorganic spin-Peierls material is known (CuGeO$_3$, with
$T_{\mathrm{sP}}=14$~K \cite{cugeo3}).

As the previous example of the VBS illustrates, allowing strong interactions from distant spins opens up new
possibilities of interesting phases to study. In fact
a rich seam of physics is revealed if start with a $d=1$
spin Hamiltonian, 
and allow the possibility of both an $XY$-like spin anisotropy component and
next-nearest neighbour interactions.  This is the content of the 
$J_{1}$-$J_{2}$ model, whose Hamiltonian is given by
\begin{equation}
\hat{H} = J_{1}\sum_{\langle i j \rangle} 
(\hat{S}_{i}^{x} \hat{S}_{j}^{x} + \hat{S}_{i}^{y} \hat{S}_{j}^{y} +
\alpha\hat{S}_{i}^{z} \hat{S}_{j}^{z})
+
J_{2}\sum_{ik} 
(\hat{S}_{i}^{x} \hat{S}_{k}^{x} + \hat{S}_{i}^{y} \hat{S}_{k}^{y} +
\alpha\hat{S}_{i}^{z} \hat{S}_{k}^{z}),
\end{equation}
where $i$ and $j$ label nearest-neighbours; $i$ and $k$ label
next-nearest neighbours and $\alpha$ is a parameter which we may vary
from 0 ($XY$ spins), through 1 (Heisenberg spins) to  large values
(resulting in Ising spins). 
This model may, in turn, be mapped onto the sine-Gordon model of
quantum field theory
\cite{sachdev,coleman} (the origin of the latter's curious name
is described in Ref~\cite{coleman}) and whose RG scaling analysis leads to the phase
diagram shown in figure~\ref{theory}(ii).

From figure~\ref{theory}(ii) we see that tuning the parameters of this model allows us to access three
ground states:
the T-L liquid; the VBS (both described above) and the Ising-N\'{e}el state. 
[It is worth noting that along the line $\alpha=1$ (corresponding to
Heisenberg spins) it can be argued that for $J_{2}/J_{1}<J_{c}$ the
system is a T-L liquid 
but
for $J_{2}/J_{1}>J_{c}$ the system adopts the VBS state (as claimed above and in section~\ref{1d})]. 
The Ising-N\'{e}el state, reached by making $\alpha>1$, with moderate
next-nearest neighbour coupling, is characterized by a spontaneously broken
symmetry at $T=0$.  Here we have magnetic LRO (not seen
in either the T-L or VBS states) involving a staggered expectation value for the
$z$-component of the spins. 
The lowest energy excitations of this phase are gapped $s=1/2$
spinons. The properties of these states are summarised in table~\ref{1dtable}.

\begin{table}
\begin{tabular}{cccc}
\hline
Phase & $T=0$ ground state & excitation & character \\
\hline
Tomonaga-Luttinger liquid & Spin disordered &$s=1/2$  spinons & gapless \\
Valence bond solid & Dimerised & spinons & gapped \\
Ising-N\'{e}el & N\'{e}el ordered &spinons & gapped \\
\hline
\end{tabular}
\caption{Phases of the $d=1$ magnet. \label{1dtable}}
\end{table}

\subsection{$d=2$: quantum criticality and spin liquids}

The ground state for the 2DQHAF is antiferromagnetically (N\'{e}el) ordered at
$T=0$ \cite{tsvelik,sachdev,manousakis}. This phase has gapless spin-wave excitations
which do not carry
a definite spin. (However, if the N\'{e}el order is aligned along $z$, the
spin waves have total $S_{z}=\pm 1$.)
As we increase the strength of the effective long-range
coupling (which we call $g$ here) another phase is realised above a
QCP at a critical
effective coupling $g_{c}$, as shown in figure~\ref{theory}(iv). This phase, which
is often called a quantum paramagnet, does not
show LRO and has a gap against $s=1$ excitations, which may be thought
of as flipped spins. 
This state can also be realised in the sort of spin model we have been
considering 
by including next-nearest neighbour interactions as we discussed
for the $d=1$ model. In fact, numerical and series expansion studies
suggest that $g_{\mathrm{c}}$ corresponds to a ratio between
next-nearest and nearest neighbours of $J_{2}/J_{1}=J_{\mathrm{c}}
\approx 0.4$. 
The arguments leading to the quantum paramagnet neglect
the contribution of Berry phases to the final configuration of the
system. If these are included a VBS state is realised with singlet
bonds lining up in columns or plaquettes on the $d=2$ square lattice.
Such a ground state is four-fold degenerate and so, as in the $d=1$
case, the VBS state that is realised must break translational
symmetry. However, the lowest lying excitation is the same as predicted
for the quantum paramagnet (a gapped $s=1$ quasiparticle), but higher
excitations will be different \cite{sachdev}.

Finally, although everything we have discussed in this section applied
for $T=0$, we may consider non-zero temperatures in this $d=2$ case,
which results in the 
phase diagram shown in
figure~\ref{theory}(iv). The curved lines represent crossovers in behavior which
allow us to identify a {\it quantum critical} region, which extends
down to $T=0$
for $g=g_{c}$. Here,
both quantum fluctuations (i.e.\ time dependence in the wavefunctions
deriving from uncertainty relations) and thermal fluctuations are
equally important and it is impossible to describe the excitations in
terms of nearly-free quasiparticles. In this region the phase of the
wavefunctions are as incoherent as Nature allows, realising an (almost)
ideal fluid. Quantum critical behaviour continues to be an area of
much experimental and theoretical interest. The properties of the
$d=2$ system are summarised in table~\ref{2dtable}

\begin{table}
\begin{tabular}{cccc}
\hline
Phase & $T=0$ ground state  & excitation & character \\
\hline
N\'{e}el ($g<g_{\mathrm{c}}$)& Staggered order  & Total $S_{z}=1$ spin waves & gapless \\
Quantum paramagnet ($g>g_{\mathrm{c}}$)& disordered & $s=1$ flipped spins & gapped \\
Valence bond solid ($g>g_{\mathrm{c}}$) & total $S=0$ &$s=1$  & gapped \\
\hline
\end{tabular}
\caption{Phases of the $d=2$ Heisenberg magnet. The inclusion of Berry
  phases in the theoretical description
leads to the VBS phase for $g>g_{\mathrm{c}}$.\label{2dtable}}
\end{table}

\begin{figure}
\begin{center}
\epsfig{file=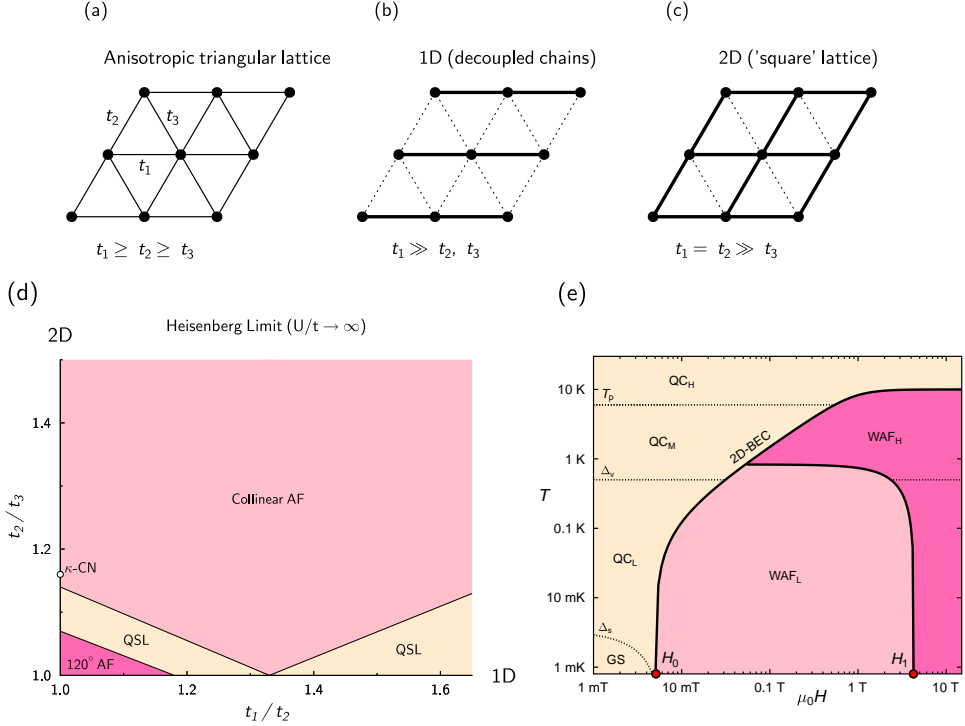,width=13cm}
\caption{(a) The triangular lattice with couplings $t_{1}$, $t_{2}$
  and $t_{3}$ as shown.  By varying the coupling we may
realize (b) the $d=1$ or (c) the $d=2$ cases. 
(d) Generalized phase diagram \cite{hauke} for the triangular
lattuce. 
(QSL=quantum spin liquid phase, $\kappa$-CN = $\kappa$-(BEDT-TTF)$_{2}$Cu$_{2}$(CN)$_{3}$.) 
(e) Phase diagram of $\kappa$-(BEDT-TTF)$_{2}$Cu$_{2}$(CN)$_{3}$
derived from $\mu^{+}$SR measurements \cite{spinliq}. \label{triangle}}

\end{center}
\end{figure}

As an example of how muons may be useful in elucidating the physics of
the exotic phases described here, we mention the case of
$\kappa$-(BEDT-TTF)$_{2}$Cu$_{2}$(CN)$_{3}$ \cite{spinliq}. 
This is a $d=2$ system based around a triangular lattice of
(antiferromagnetically coupled) $s=1/2$ spins, whose low energy physics
provides
an example of a frustrated, non-collinear antiferromagnet.
In fact, the triangular lattice, shown in figure~\ref{triangle}(a)
may be viewed as a meeting point of the $d=1$ and $d=2$ cases. 
We may recover those cases by tuning the coupling constants
to form either well coupled $d=1$ chains (e.g.\ by taking $t_{1}\gg t_{2},
t_{3}$) [figure~\ref{triangle}(b)]
or, with the choice $t_{1}=t_{2} \gg t_{3}$, forming a lattice which
is topologically equivalent to the $d=2$ square lattice case [figure~\ref{triangle}(c)]. 
 A general phase diagram for the triangular
lattice, 
derived from calculations using 
modified spin wave theory  \cite{hauke} and assuming perfectly
localized Heisenberg spins,
is shown in Fig.~\ref{triangle}(d). The $d=1$ case is realized far along the horizontal axis, 
where the T-L ground state is identified as one example of
a quantum spin liquid (QSL) state (here used as a general term used to describe
a $T=0$ ground state without LRO, of which there are potentially very
many varieties). The $d=2$ case is found far
along the vertical axis showing the expected $T=0$ collinear
(N\'{e}el) LRO. Of
note in figure~\ref{triangle}(d) is the isotropically coupled case
($t_{1}=t_{2}=t_{3}$) where the classical solution of
antiferromagnetic LRO with a 120$^{\circ}$ spin structure is realised,
albeit with moments much reduced by quantum fluctuation. This neighbours a QSL state
for the case where the couplings are slightly anisotropic. 

Returning to the case of $\kappa$-(BEDT-TTF)$_{2}$Cu$_{2}$(CN)$_{3}$,
we see that its couplings appear to place it on the border of the QSL and N\'{e}el phases
in figure~\ref{triangle}(d). In fact, a more realistic description of
the material is acheived by assuming that $U/t$ (the ratio of the electronic
correlation energy to the bandwidth) is finite. 
This, which amounts physically to a relaxation of the assumption
that the relevant electronic degrees of freedom are perfect Heisenberg spins (or
so-called ``quantum tops'') is expected to
destabilise phases showing LRO near the origin in figure~\ref{triangle}(d),
 enlarging the QSL phase to such an extent that 
 $\kappa$-(BEDT-TTF)$_{2}$Cu$_{2}$(CN)$_{3}$ should be expected to
realize a QSL ground state. 
In fact, the $\mu^{+}$SR measurements \cite{spinliq}
suggest that the material
realizes a $Z_{2}$ QSL ground state showing {\it
  topological order} \cite{sachdev}, with total $S=0$ and
$s=1/2$ spinon excitations with a very small gap against their
creation. (The ``$Z_{2}$'' assignment 
tells us that the system is described by a topological {\it gauge
  field theory} \cite{coleman} somewhat similar to those used to
describe fractional quantum Hall fluids,
whose description is locally invariant under the $Z_{2}$ group of transformations of the fields.)
There is also evidence for another type of gapped excitation in
this system, which is a point-like vortex known as a {\it vison}. For 
$\kappa$-(BEDT-TTF)$_{2}$Cu$_{2}$(CN)$_{3}$ muon measurements \cite{spinliq} not only 
suggest this complex excitation spectrum, but the application of
magnetic fields allows the system to be driven into other phases
showing LRO with extremely small magnetic moments, leading to the
phase
diagram shown in figure~\ref{triangle}(e). The exquisite sensitivity of
$\mu^{+}$SR and the potential to apply it to samples which are smaller
than those required for inelastic neutron measurements have made it
particularly useful in this context. 

\section{Conclusion}

In the short term muons will continue to allow us to probe low-moment LRO 
and aspects of the excitation spectrum in low-dimensional
magnetic systems based on molecular building blocks. In addition to
the
cases outlined here there are many other avenues for future study
in this field, including the investigation of $d=0$ systems such as 
those based on dimers or on molecular nanomagnets formed from clusters containing
large numbers of strongly coupled spins. Also of interest are
spin-ladder systems which interpolate between the harsh extremes of
$d=1$ and $d=2$. In the future, the possibility of being able to work in more
extreme conditions involving large applied magnetic fields and
pressures will open
out more of the phase diagrams of these systems. At the same time, advances in
materials synthesis will allow us to approach the platonic
ideals of the models described in this Comment.

\ack
We would like to express particular thanks to J L Manson, J A
Schlueter and C P Landee for their long-term collaboration on this work. We are also most
grateful to all of our collaborators and students and to the
staff of ISIS and PSI muon facilities. 
We thank P Hauke for providing modified spin wave theory results and P A Goddard and F Xiao for useful discussions 
regarding this Comment.
TL is supported by EPSRC.


\begin{thebibliography}{10}
\bibitem{note}
{We use the
single-$J$ convention here where the energy gap between the singlet and
triplet level of a two spin system with Hamiltonian $\hat{H} = J
\boldsymbol{S}_{i}\cdot\boldsymbol{S}_{j}$
is given by $J$. }
\end{thebibliography}
\end{document}